\documentclass[11pt]{article}

\usepackage[tmargin=1.5cm,bmargin=2.5cm,lmargin=2.5cm,rmargin=2.5cm]{geometry}
\usepackage{amsmath,color,graphicx,amssymb,bm,mathrsfs,amsthm,cite,authblk}

\usepackage{sectsty}
\sectionfont{\large\selectfont}

\usepackage{tgtermes}

\newtheorem{corollary}{\bf Corollary}[section]
\newtheorem{lemma}{\bf Lemma}[section]
\newtheorem{proposition}{\bf Proposition}[section]

\newcommand{\REV}[1]{\textcolor{black}{#1}}

\newcommand{\Gr}{\mathit{Gr}}
\newcommand{\Rey}{\mathit{Re}}

\newcommand{\tom}{F}
\newcommand{\rem}[1]{}

\newcommand{\bel}{\begin{equation}\label}
\newcommand{\ee}{\end{equation}}
\newcommand{\beq}{\begin{eqnarray}} 
\newcommand{\eeq}{\end{eqnarray}} 
\newcommand{\bc}{\begin{center}} 
\newcommand{\ec}{\end{center}} 
\newcommand{\ben}{\begin{enumerate}}
\newcommand{\een}{\end{enumerate}}
\newcommand{\bit}{\begin{itemize}}
\newcommand{\eit}{\end{itemize}}
\newcommand{\I}{\int_{\Omega}}

\newcommand\shalf{\ensuremath{{\scriptstyle\frac{1}{2}}}}

\newcommand{\bom}{\mbox{\boldmath$\omega$}}

\newcommand{\bu}{\mbox{\boldmath$u$}}

\newcommand{\bx}{\mbox{\boldmath$x$}}

\title{\sc Dynamics of vorticity moments in shell models of turbulence: A comparison with the Navier--Stokes equations}

\author[1]{John D. Gibbon\thanks{j.d.gibbon@ic.ac.uk}}
\author[2]{Dario Vincenzi\thanks{dario.vincenzi@univ-cotedazur.fr}}
\affil[1]{\it\normalsize Department of Mathematics, Imperial College London, London SW7 2AZ, UK}
\affil[2]{\it\normalsize Universit\'e C\^ote d'Azur, CNRS, LJAD, 06100, Nice, France}

\begin{document}

\maketitle

\begin{abstract}
Shell models allow much greater scale separations than those presently achievable with direct numerical simulations of the Navier--Stokes equations. Consequently, they are an invaluable tool for testing new concepts and ideas in the theory of fully developed turbulence. They also successfully display energy cascades and intermittency in homogeneous and isotropic turbulent flows. Moreover, they are also of great interest to mathematical analysts because, while retaining some of the key features of the Euler and the Navier--Stokes equations, they are much more tractable. A comparison of the mathematical properties of shell models and of the three-dimensional Navier--Stokes equations is therefore essential in understanding the correspondence between the two systems. Here we focus on the temporal evolution of the moments, or $L^{2m}$-norms, of the vorticity. Specifically, differential inequalities for the moments of the vorticity in shell models are derived. The contribution of the nonlinear term turns out to be much weaker than its equivalent for the three-dimensional Navier--Stokes equations. Consequently, pointwise-in-time estimates are shown to exist for the vorticity moments for shell models of any order. This result is also recovered via a high-low frequency slaving argument that highlights the scaling relations between vorticity moments of different orders. Finally, it is shown that the estimates for shell models formally correspond to those for the Navier--Stokes equations \REV{`on a point'}.\end{abstract}

\section{Introduction}
Faced with the difficulty of the three-dimensional Navier--Stokes regularity problem, and in an effort to
gain analytical insight into it, in past years mathe\-ma\-ti\-cians have turned to hydrodynamic models that capture the phenomenology of isotropic turbulence. Shell models are a particular example. They are infinite systems of nonlinear ordinary differential equations for a set of velocity variables which can be interpreted as the Fourier modes of the velocity field \cite{frisch1995,bjpv98,biferale2003,psf13,bt23}.  They also enjoy the advantage of being more tractable than the Navier--Stokes and the Euler equations (see Ref.~\cite{cdf23}, for a recent review). 
The nonlinear interactions between the velocity variables are truncated to the nearest and next-to-nearest modes, which considerably simplifies both the mathematical analysis and numerical simulations. 
Shell models indeed  allow much higher scale separation and greater statistical convergence than those presently achievable with direct numerical simulations of the Navier--Stokes equations~\cite{dewit2024}. The global regularity of solutions of shell models with deterministic forcing was established by Constantin {\it et al.} \cite{clt06,clt07a}. Upper and lower bounds for the Hausdorff dimension of the global attractor were provided, and it was proved that shell models possess a finite-dimensional inertial manifold. For a large-scale forcing, it was also shown that the velocity modes decay exponentially at large wave-numbers, which implies the existence of a viscous dissipation range. In the inviscid case,  the same authors established the global existence of weak solutions and their uniqueness for short times \cite{clt07b}. Moreover, they provided the shell-model analogues of Onsager's conjecture and the Beale--Kato--Majda criterion for the Euler equations \cite{clt07b}. The latter criterion was used in Ref.~\cite{mailybaev12} to show the existence of blowup in the inviscid Sabra shell model. Rigorous results for the case of stochastic forcing were obtained in Refs.~\cite{bbbf06,bf12,bgas16,bhr16}.

\par\smallskip
A comparison between shell models and the three-dimensional Navier--Stokes equations was presented in Ref.~\cite{vg21}.  It was shown, in particular, that the estimates for the time averages of the Sobolev norms of the velocity and their moments display a milder Reynolds number dependence for shell models than for the three-dimensional Navier--Stokes equations. Rather, the estimates for the two systems become equivalent if the Navier--Stokes equations are supplemented with an `intermittency-suppression' assumption, which consists in requiring that the ratio of the $L^\infty$- and $L^2$-norms of the velocity gradient is independent of the Reynolds number.
Here we continue the comparison between the two systems by studying the temporal dynamics of the moments of vorticity. This provides further insight into the similarities and the differences between shell models and the Navier--Stokes equations.

\par\smallskip
In \S\ref{sect:NS}, we summarize the known results on the three-dimensional Navier--Stokes equations. We introduce a set of rescaled vorticity norms and recall the scaling of their long-time averages with the Reynolds number. Previous numerical simulations suggest that the higher vorticity moments are controlled by a power of the enstrophy weaker than the square root. Such a numerical observation can be made rigorous via a high--low frequency slaving assumption. When used in the differential inequalities for the rescaled vorticity norms, this fact yields a depletion of the vortex stretching term and ultimately a conditional proof of regularity. In \S\ref{sect:shell-def}, we introduce the Sabra shell model and the analogue of the rescaled vorticity norms in this model. In \S\ref{sect:shell-diff-ineq}, we derive the differential inequalities that govern the time evolution of the rescaled vorticity moments. One of the main consequences of such inequalities is the existence of absorbing balls for moments of any order. Finally, in \S\ref{sect:discussion}, the shell-model results are compared with their Navier--Stokes counterpart. A formal correspondence between shell models and the Navier--Stokes equations `on a point' is also discussed.

\section{Summary of Navier--Stokes results}\label{sect:NS}

\subsection{A sequence of rescaled vorticity norms}\label{subsect:vort}

Consider the incompressible, forced, three-dimensional Navier--Stokes equations for the velocity field $\bu(\bx,t)$ on the periodic domain $\mathscr{V}=[0,L]^3$:
\begin{equation}
\partial_t\bu + \bu\cdot\nabla\bu = -\nabla p +\nu\Delta\bu + \bm f(\bx), \qquad \operatorname{div}\bu=0,
\end{equation}
where $\nu$ is the kinematic viscosity and $p$ the pressure. The forcing $\bm f(\bx)$ is time-independent, divergence-free, mean-zero, and narrow-band. Moreover, its longest length scale is taken equal to $L$ for convenience. The Grashof and Reynolds numbers are defined as 
\begin{equation}
\Gr = L^{3/2}\Vert\bm f\Vert_2 \nu^{-2}, \qquad \Rey = UL \nu^{-1},
\end{equation}
where $U^2=L^{-3}\langle\Vert\bu\Vert_2^2\rangle_t$ with 
\begin{equation}
\REV{\langle\cdot\rangle_t= \frac{1}{t}\int_0^t \cdot\,dt}
\end{equation}
denoting the time average up to time $t>0$. 
While the energy $E=\Vert\bu\Vert_2^2$ is uniformly bounded in time in this setting, strong solutions of the three-dimensional Navier--Stokes equations are known to exist globally in time only for small initial data\,; otherwise their existence can be guaranteed only up to a finite time \cite{dg95,foias2001,ld08,doering2009,robinson2016,robinson2020,vicol2022}. The main result for the three-dimensional Navier--Stokes equations has been the establishment of Leray's weak solutions \cite{Leray1934}. The next big advance was the result of Foias, Guillop\'e and Temam \cite{fgt1981,foias2001}, which essentially shows the boundedness of a hierarchy of time averages
\bel{fgt1}
\REV{\Big<\|\nabla^{n}\bu\|_{2}^{\frac{2}{2n-1}}\Big>_{t} < \infty\,.}
\ee
In Ref.~\cite{gibbon2023} this has been generalized from $n$ derivatives in $L^{2}(\mathscr{V})$ to $L^{2m}(\mathscr{V})$ in $d=2,3$ spatial dimensions on the domain $\mathscr{V}=[0,\,L]^{d}$\,:
\bel{ta23}
\REV{\nu^{-(4-d)\alpha_{n,m,d}}L^{4-d}\left<\|\nabla^{n}\bu\|_{2m}^{(4-d)\alpha_{n,m,d}} \right>_{t} \leq c_{n,m,d}\Rey^{3} +O\big(t^{-1}\big)}
\ee
with
\bel{alphadef}
\alpha_{n,m,d}=\frac{2m}{2m(n+1)-d}\,.
\ee
$n,~m$ satisfy $1\leq m \leq \infty$ and $n\geq 1$, \REV{while $c_{n,m,d}$ are dimensionless constants}. The exponent $\alpha_{n,m,d}$ arises naturally from the scaling properties of the Navier--Stokes equations and is also consistent with the application of Sobolev and H\"older inequalities \cite{gibbon19,gibbon2020epl}.

The time averages in \eqref{ta23} encapsulate all the weak solution results for the Navier--Stokes equations \cite{gibbon2023}. However, the aim of this paper is to consider higher moments of the vorticity so instead of considering arbitrarily high derivatives we remain\footnote{While $\nabla\bu$ and $\bom$ are synonymous in $L^{2}(\mathscr{V})$ they are not in $L^{p}(\mathscr{V})$\,: in fact, $\|\bom\|_{p}\leq \|\nabla\bu\|_{p}$ for $2 \leq p\leq \infty$ whereas $\|\nabla\bu\|_{p}\leq c_{p}\|\bom\|_{p}$ for $2 \leq p < \infty$. The inequality needs a logarithmic correction on the RHS because at $p=\infty$ the constant $c_{p} $ blows up. We refer to the $L^p$-norms $\Vert\bom\Vert_p$ as (spatial) moments of the vorticity; these should not be confused with the statistical moments of $\vert\bom\vert$ that are commonly studied in the statistical theory of turbulence \cite{frisch1995}.} with $n=1$.   Specifically, we use the spatial moments of the vorticity field $\bom=\operatorname{curl}\bu$ with the dimension of a frequency
\begin{equation}
\Omega_m(t)=L^{-3/2m}\Vert\bom\Vert_{2m}\,, \qquad 1\leqslant m\leqslant \infty\,,
\end{equation}
where $m$ is not necessarily integer. Moreover, $\Omega_m(t)$ are ordered as follows for all $t$\,: 
\begin{equation}
\label{eq:omega-order}
\Omega_1(t) \leqslant \cdots \leqslant \Omega_{m}(t)\leqslant\Omega_{m+1}(t)\leqslant \cdots \leqslant \Omega_\infty(t)\,.
\end{equation} 
Firstly, we define $\varpi_0=\nu L^{-2}$ as the basic frequency associated with the domain $\mathscr{V}$. Then we note that $\nu^{-1}L^{1/\alpha_{1,m,3}} = \varpi_{0}^{-1}L^{-3/2m}$. Thus, when $\nabla\bu$ is replaced by $\bom$, the equivalent of \REV{$\nu^{-\alpha_{1,m,3}}L \|\nabla\bu\|_{2m}^{\alpha_{1,m,3}}$} in (\ref{ta23}) with $n=1$ when $d=3$ is
\bel{Dmdef}
D_m(t)=\left[\varpi_0^{-1}\Omega_m(t)\right]^{\alpha_{m}}, \qquad 1\leqslant m\leqslant \infty\,.
\ee
The nomenclature $D_{m}$ in (\ref{Dmdef}) had previously been introduced in \cite{gibbon12cms,donzis13}. 
With $n=1$, $\alpha_{1,m,3} = \frac{2m}{4m-3} \equiv \alpha_{m}$ with the $\alpha_{m}$ having been introduced 
in two earlier papers \cite{gibbon12cms,gibbon19}, where it was also shown that\footnote{To be precise, the estimate of the time average of $D_m$ is 
\begin{equation}
\langle D_m\rangle_t \leqslant \hat{c} \Rey^2 + c \Rey^3 + O\big(t^{-1}\big),\qquad 1\leqslant  m\leqslant \infty
\label{eq:full-bound}
\end{equation}
\REV{with $\hat{c}$ a dimensionless constant.}
Here the focus is on the large-$\Rey$ regime. Based on the bound 
\begin{equation}
\Gr \leqslant c' \Rey + c'' \Rey^2 + O\big(t^{-1}\big)
\end{equation}
\REV{with $c'$ and $c''$ dimensionless constants depending on the `forcing shape' alone \cite{df02},} the large-$\Rey$ regime
is obtained for $\Gr\gg 1$. Therefore, throughout the paper we will assume $\Gr\gg 1$ and ignore the $\Rey^2$  term in \eqref{eq:full-bound}.
}
 
\begin{equation}
\label{eq:Dm_avg}
\langle D_m\rangle_t \leqslant c\,\Rey^3 + O\big(t^{-1}\big),\qquad 1\leqslant  m\leqslant \infty
\end{equation}
\REV{where the uniform constant $c$ can be obtained by maximizing $c_{1,m,3}$ with respect to $m$ \cite{gibbon12cms}.} For a variety of forcings and initial conditions, numerical simulations of high-$\Rey$ turbulence in a periodic box have shown that the bound in \eqref{eq:Dm_avg} is saturated if $m=1$ \cite{donzis13}. However,  when $m>1$ the time average of  $D_m$ has been found to grow more slowly than $\Rey^3$ \cite{donzis13}. Two possible reasons may be at the origin of this different scaling: 1) higher Reynolds numbers may be required to observe the $\Rey^3$ scaling for $m>1$ or 2) the bound may be saturated for all $m$ only in the case of some specific initial conditions that are irrelevant to the turbulent regime. 

\subsection{High--low frequency relation and  the ordering of the $D_m$}
\label{sect:high-low}

In principle, the $D_m$ do not satisfy any natural ordering, because while $\Omega_m$ is an increasing function of $m$, the exponent $\alpha_m$ decreases with $m$. Nevertheless, numerical simulations show that the $D_m$ are ordered on a descending scale such that $D_{m+1} < D_{m}$ for $1\leqslant m\leqslant\infty$ and, remarkably, with a very large separation between $D_m$ of different orders, especially for the lowest $m$\cite{donzis13} (for a study of the $D_m$ in the three-dimensional Euler equations, see Ref.~\cite{kerr2013}). As a matter of fact, the numerical data turn out to be accurately described by a relation of the form\footnote{The $m$-subscript on $\mu$ in the subscript of $A_{m,\mu}$ has been suppressed.}
\begin{equation}
\REV{D_m = b_m D_1^{A_{m,\mu}(\tau)}, \qquad 2\leqslant m\leqslant \infty\,,}
\label{eq:Am}
\end{equation}
where $\tau=\varpi_0 t$ is a dimensionless time, \REV{$b_m$ are dimensionless constants,} 
\begin{equation}
\REV{A_{m,\mu}=\dfrac{(m-1)\mu_m(\tau)+1}{4m-3}\,,}
\end{equation}
and \REV{$\mu_m(\tau)$} lies in the interval $1.15\leqslant\mu_m(\tau)\leqslant 1.5$ \cite{gibbon14}. This numerical observation can be derived analytically under a high--low frequency slaving assumption \cite{gibbon2016ima}. Here we present an alternative derivation which includes some variations on that given in Ref.~\cite{gibbon2016ima}. Let us define 
\bel{f1}
F_{m} = \varpi_{0}^{-1}\Omega_m\,.
\ee
From \eqref{eq:omega-order}, the  dimensionless frequencies $\tom_{m}$ must satisfy Holder's inequality:
\bel{Hin1}
\tom_{1} \leqslant  \ldots \leqslant  \tom_{m} \leqslant  \tom_{m+1} \leqslant  \ldots
\ee
There has been a long-standing folk-belief that, in nonlinear systems, `high modes' might depend on `low modes'. The idea of the inertial manifold in the 1980s was a rigorous version of this \cite{fst88,ft91,foias2001}. This worked for the one-dimensional Kuramoto--Sivashinsky equation \cite{foias1988pla} and, more recently, for shell models \cite{clt06}. Following this idea, we start with the \textit{ansatz}
\bel{f2}
\tom_{m} = F_{m}(\tom_{1},\, \tom_{2},\, \ldots\,,\, \tom_{n})\,,
\ee
for some $n > 1$.  We need more than this as (\ref{f2}) must be constrained and shaped by both (\ref{Hin1}) and the triangular version of H\"older's inequality (see the appendix of Ref.~\cite{gibbon2016ima})
\bel{Hin2}
\left(\frac{\tom_{m}}{\tom_{1}}\right)^{m^{2}} \leqslant  \left(\frac{\tom_{m+1}}{\tom_{1}}\right)^{m^2 - 1}\,.
\ee
Firstly, it is well known that the existence and uniqueness of solutions of the three-dimensional Navier--Stokes equations depend entirely on the boundedness of \REV{$\Vert\nabla\bu\Vert_2^2$}, which is proportional to $\tom_{1}$\,: thus we treat $F_{1}$ as the main variable on which all others depend. Secondly, the form of the two inequality constraints (\ref{Hin1}) and (\ref{Hin2}) suggests that we simplify the dependency of $\tom_{m}$ to the first frequency $\tom_{1}$ such that our \textit{ansatz} in (\ref{f2}) is reduced to
\bel{f3}
\tom_{m} = \tom_{1}\Phi_{m}(\tom_{1})\,.
\ee
(\ref{Hin1})  and (\ref{Hin2}) then demand that $\Phi_{m}$ must satisfy both 
\bel{f4}
\Phi_{m+1} \geqslant  \Phi_{m}^{\frac{m^{2}}{m^{2}-1}} \qquad\mbox{and}\qquad\Phi_{m} \geqslant  1\,.
\ee
The first inequality in (\ref{f4}) can be further simplified by the substitution
\bel{f5a}
\Phi_{m} = h_{m}^{\left(\frac{m-1}{m}\right)\beta}
\ee
for a smooth, arbitrary function $\beta(\tau) > 0$. This reduces both inequalities in (\ref{f4}) to the monotonically increasing sequence
\bel{f5b}
1 \leqslant  h_{m}\left(\tom_{1}\right) \leqslant   h_{m+1}\left(\tom_{1}\right)
\ee
and transforms \eqref{f3} into
\bel{f5c}
\tom_{m} = \tom_{1}\left[h_{m}\left(\tom_{1}\right)\right]^{\left(\frac{m-1}{m}\right)\beta}\,.
\ee
The function $h_{m}\left(\tom_{1}\right)$ is too general to be of use until we recall that it must be subject to one further 
constraint, which is the fact that the time averages of $F_{m}^{\alpha_{m}}$ must be bounded. Thus the $h_{m}$ cannot be 
any stronger than a power law in $\tom_{1}$ which we write as
\bel{f6a}
h_{m}\left(\tom_{1}\right) = \tom_{1}^{\gamma_{m}}
\ee
where $\gamma_{m} > 0$. 
With the definitions
\begin{align}\label{6c}
\begin{split}
\mu_{m}(\tau) &= 1 + \gamma_{m}\beta(\tau)\qquad\mbox{with}\qquad \mu_{m} \leqslant  \mu_{m+1}\,,
\end{split}
\end{align}
(\ref{f5c}) transforms to
\bel{f5e}
F_{m} = F_{1}^{\frac{(m-1)\mu_{m}(\tau) + 1}{m}}\,,
\ee
which is \eqref{eq:Am} \REV{up to a multiplicative constant}.
Note that the exponent is positive and $\mu_{m} \geqslant  1$. To be sure that, for weak solutions, we can use the finiteness of $\big<F_{1}^{2}\big>_t$  in order to ensure the finiteness of $\left<F_{m}^{\alpha_m}\right>_t$, we must also have
\begin{equation}
\alpha_m[(m-1)\mu_m(\tau)+1]\leqslant 2m\,,
\label{eq:condition-lambda}
\end{equation}
whence
\bel{f5f}
1 \leqslant  \mu_{m} \leqslant  4\,.
\ee
For full regularity, however, $\mu_m$ would have to be less than 2 (see Ref.~\cite{gibbon19} and \S\ref{sect:diff-ineq} below).

\subsection{Differential inequalities for $D_1$}
\label{sect:diff-ineq}

For a single-scale forcing ($\Vert\nabla\bm f\Vert_2 \approx L^{-1}\Vert\bm f\Vert_2$) and under the assumption that strong solutions exist, $D_1$ satisfies the differential inequality
\begin{equation}
\frac{\dot{D}_1 }{2\varpi_0}\leqslant -\frac{\nu^2 L}{4E}\,D_1^2 + c_1 D_1^{3}+ \Gr D_1^{1/2},
\label{eq:ineq-D1-NS}
\end{equation} 
where the dot denotes differentiation with respect to time \cite{gibbon14}. Inequality  \eqref{eq:ineq-D1-NS} holds for strong solutions and hence for short times or small initial data. Indeed, although the energy is always bounded, the vortex stretching term is cubic and, for large values of $D_1$, cannot be countered by the viscous term, which is only quadratic. 
Some progress can be made by invoking the nonlinear depletion in \eqref{eq:Am}.
Using \eqref{eq:Am} in the estimate of the vortex-stretching term yields\footnote{\REV{Note that $c_1$ and $c_m$ in \eqref{eq:ineq-D1-NS} and \eqref{eq:Dm-lambda} are not related to the constants $c_{n,m,d}$ in \S~\ref{subsect:vort}.}} \cite{gibbon14, gibbon2016ima}
\begin{equation}
\frac{\dot{D}_1 }{2\varpi_0} \leqslant -\frac{\nu^2 L}{4E}\,D_1^2 + c_m D_1^{1+\mu_m/2} + \Gr\,D_1^{1/2}.
\label{eq:Dm-lambda}
\end{equation}
Since $1\leqslant \mu_m\leqslant 4$, the vortex stretching term is no stronger than cubic. If in addition $1\leqslant\mu_m < 2$, as is observed in numerical simulations \cite{gibbon14}, the dissipative term prevails for large values of $D_1$. Consequently, it can be shown that there exist an absorbing ball for $D_1$ \REV{such that, for $\Gr\gg 1$,
\begin{equation}
\label{eq:ns-pointwise}
\varlimsup_{t\to\infty} D_1 \leqslant \tilde{c}_m \Gr^{4/(2-\mu_m)},
\end{equation}
where $\tilde{c}_m$ is a dimensionless constant,}
and a global attractor for solutions of the Navier--Stokes equations \cite{gibbon14,gibbon2016ima}. This remains, however, a conditional regularity result.
Finally, analogous differential inequalities exist for higher-order $D_m$ \cite{gibbon14,gibbon2016ima}.

\section{Shell-model: definitions and preliminary properties}\label{sect:shell-def}

Having recalled the main results on the vorticity moments of Navier--Stokes solutions in \S\ref{sect:NS}, we now turn to the main goal of this paper, namely the dynamics of vorticity moments in shell models.
\par\smallskip
In the Sabra shell model \cite{sabra}, the complex variables $u_j$ satisfy the system of ordinary differential equations\footnote{The results of this paper are also valid for the GOY model \cite{g73,yo87}.}
\begin{equation}
  \dot{u}^{\vphantom{*}}_{j} =
  \mathrm{i}(a_1 k^{\vphantom{*}}_{j+1} u^*_{j+1} u^{\vphantom{*}}_{j+2}
  +a_2 k^{\vphantom{*}}_j u^{\vphantom{*}}_{j+1} u^*_{j-1}
  -a_3 k^{\vphantom{*}}_{j-1} u^{\vphantom{*}}_{j-1} u^{\vphantom{*}}_{j-2})
  -\nu k_j^2 u^{\vphantom{*}}_j + f^{\vphantom{*}}_{j}, \qquad j=1,2,3,\dots,
\label{eq:shell}
\end{equation}
where $*$ denotes complex conjugation, $\nu$ is the kinematic viscosity, $f_j$ are the forcing variables, and $k_j=k_0\lambda^j$ with $k_0>0$ and $\lambda>1$. The `boundary conditions' are $u_0=u_{-1}=0$, while $a_1$, $a_2$, $a_3$ are real. \REV{In the following, the constants $a_i$ with $i=1,2,3$ will be reserved for the coefficients of the shell model.}
The condition $a_1 + a_2 + a_3 = 0$ ensures that the kinetic energy\footnote{For ease of comparison, we use the same notation for quantities that have the same meaning in the shell model and the Navier--Stokes equations. \REV{The values of the dimensionless constants obviously differ from those that appear in the corresponding Navier--Stokes estimates.}} 
\begin{equation}
E=\sum_{j=1}^\infty \vert u_j\vert^{2}
\end{equation}
is preserved  when the system is inviscid ($\nu=0$) and unforced ($f_j=0$ for all $j$). We assume that the initial energy is finite.

\REV{We consider a forcing of the form
\begin{equation}
  f_j=
  \begin{cases}
    0,  & j < j_f,
    \\
    F\phi_j, &  j \geqslant j_f,
  \end{cases}
\end{equation}
where $F$ is a complex constant and $\phi_j$ is time independent and such that 
\begin{equation}
\label{eq:coni-phi}
\sum_{j=1}^\infty k_j^2\vert\phi_j\vert^2 < \infty
\end{equation}
and 
\begin{equation}
\label{eq:norm-phi}
\sum_{j=1}^\infty \vert\phi_j\vert^2 =1.
\end{equation}
Therefore, $k_f=k_0\lambda^{j_f}$ is the smallest wavenumber of the forcing. Obviously, this definition includes forcings that act on a finite set of wavenumbers, which are the most common in numerical simulations of shell models.} In the following, we shall refer to $\phi_j$ as the `shape' of the forcing.
As recalled in the introduction, the solutions of the shell model are regular under these assumptions (and in fact for more general forcings) \cite{clt06}. In particular, the energy is uniformly bounded in time and satisfies the point-wise estimate
\REV{
\begin{equation}
\varlimsup_{t\to\infty} E(t) \leqslant \nu\varpi_0 \rho_f^4 \Gr ^2,
\label{eq:point-wise-energy}
\end{equation}
where now $\varpi_0=\nu k_f^2$, $\rho_f=k_f/k_1$ is the aspect ratio of the forcing, and the Grashof number is
\begin{equation}
\Gr  = \frac{\vert F\vert}{\nu^{2} k_f^{3}}\,.
\end{equation}
This can be proved by multiplying \eqref{eq:shell} by $ u_j^*$ and the complex conjugate of \eqref{eq:shell} by  $ u_j$, adding the two resulting equations, and summing over $j$ to find
\begin{equation}
\dot{E} = -2\nu \sum_{j=1}^\infty k_j^2 \vert u_j\vert^2 + \sum_{j=1}^\infty (f_j u_j^* + f_j^* u_j ).
\end{equation}
The nonlinear term does not contribute because we have taken $a_1+a_2+a_3=0$. We now apply the shell-model counterpart of the Poincar\'e equation to the dissipative term and the Cauchy-Schwarz inequality to the forcing term and use \eqref{eq:norm-phi}\,:
\begin{equation}
\dot{E} \leqslant -2\nu k_1^2 E + 2 \vert F\vert E^{1/2}.
\end{equation}
Applying Young's inequality to the forcing term yields
\begin{equation}
\dot{E} \leqslant - \nu k_1^2 E +\frac{\vert F\vert^2}{\nu k_1^{2}}\,.
\end{equation}
Finally, \eqref{eq:point-wise-energy} follows from using Gr\"ownall's inequality.
The bound in \eqref{eq:point-wise-energy} will be instrumental in proving the point-wise boundedness of $D_m$. In the derivation,  we have used the condition $a_1+a_2+a_3=0$, which ensures the conservation of energy in the inviscid and unforced case.
  Therefore, even though in this study the coefficients of the shell model only determine the values of some multiplicative constants, the condition $a_1+a_2+a_3=0$ is essential.}

\begin{figure}[t]
\centering
\includegraphics[width=0.517\textwidth]{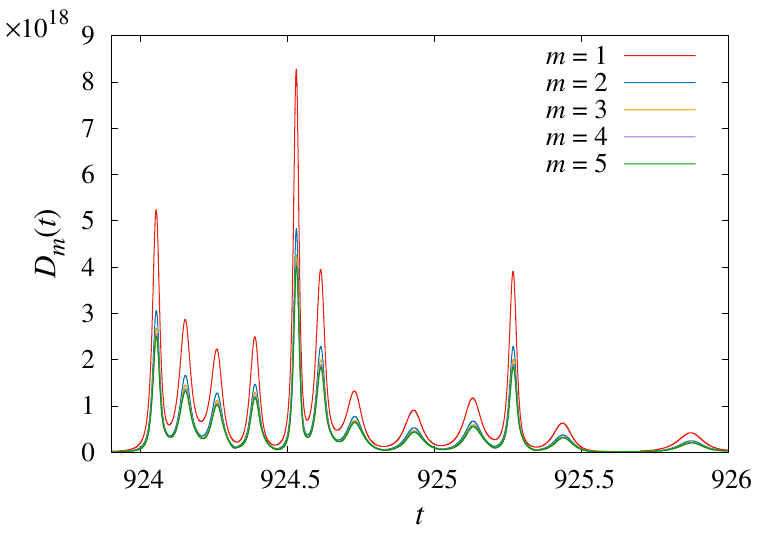}\hfill%
\includegraphics[width=0.483\textwidth]{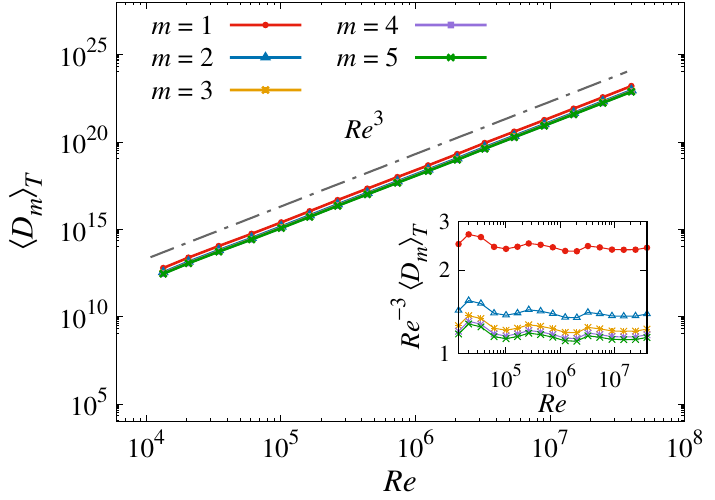}
\caption{Left: A representative series of spiky events in the time evolution of $D_m(t)$ for $m=1,\dots,5$ and $\Rey\approx 7.4\times 10^5$. Right: Long-time average of $D_m$ vs $\mathit{Re}$ for $m=1,\dots,5$. The inset shows the same time averages rescaled by $\Rey^3$.
In the simulations, $k_0=2^{-4}$, $\lambda=2$, $a_1=1$, $a_2=a_3=-1/2$,  $3\times 10^{-4}\leqslant \nu\leqslant 10^{-7}$, and $f_j=F\delta_{j,1}$ with $F=5\times 10^{-3}(1+\mathrm{i})$.
The shell model is truncated to $N=27$ shells and is integrated numerically via an Adams--Bashforth scheme \cite{pisarenko1993}.}
\label{fig:Dm}
\end{figure}

The Reynolds number is defined as
\begin{equation}
\Rey = \frac{\langle E\rangle_t^{1/2}}{\nu k_f}\,.
\end{equation}
The Doering and Foias relation between $\Gr $ and $\Rey $ has a shell-model counterpart \cite{vg21}
\begin{equation}
\label{eq:Gr-Re_shell}
\Gr  \leqslant c' \Rey  + c'' \Rey ^2,
\end{equation}
where $c'$ and $c''$ are positive constant that only depend on the `shape' of the forcing \REV{(see Ref.~\cite{vg21} for their explicit expressions).}
For the shell model, the analogues of the vorticity moments are
\begin{equation}
\Omega_m=\bigg(\sum_{j=1}^\infty k_j^{2m} \vert u_j\vert^{2m}\bigg)^{1/2m}, \qquad 1 \leqslant m < \infty\,,
\end{equation}
and
\begin{equation}
\Omega_\infty = \sup_{1\leqslant j} k_j \vert u_j\vert\,,
\end{equation}
which are nothing but the $\ell^{2m}$- and $\ell^{\infty}$-norms of the sequence $(k_j u_j)_{j\geqslant 1}$ and can be regarded as the shell-model analogues of the usual norms in the Sobolev spaces $W^{1,2m}(\mathscr{V})$ and $W^{1,\infty}(\mathscr{V})$ (note that, in shell models, there is no distinction between the moments of vorticity and those of the velocity gradient).
Contrary to the Navier--Stokes equations, the $\Omega_m$ are ordered on a descending scale:
\begin{equation}
\Omega_{m+1}\leqslant \Omega_m\,, \qquad 1 \leqslant m \leqslant \infty\,.
\end{equation}
Moreover, the appropriate rescaled moments are now \cite{vg21}
\begin{equation}
D_{m} = \left(\varpi_{0}^{-1}\Omega_m\right)^2, \qquad 1 \leqslant m \leqslant \infty\,.
\end{equation}
We observe here a first difference with the Navier--Stokes equations, because now the $D_m$ are ordered on a descending scale by definition (figure~\ref{fig:Dm}, left panel)\,:
\begin{equation}
\label{eq:order}
D_{m+1}\leqslant D_m, \qquad 1 \leqslant m \leqslant \infty\,.
\end{equation}
For the turbulent regime of the Navier--Stokes equations, this ordering was a numerical observation but did not hold a priori \cite{donzis13}.

The time average of $D_m$ satisfies an analogous bound to that valid for the Navier--Stokes equations\,:
\begin{equation}
\label{eq:time_average}
\langle D_{m}\rangle_t \leqslant C\, \textit{Re}^3 + O(t^{-1})\,,  \qquad 1 \leqslant m \leqslant \infty\,,
\end{equation}
\REV{where the dimensionless constant $C$ depends on the parameters of the shell model ($a_1$, $a_2$, $a_3$, and $\lambda$) as well as on the shape of the forcing \cite{vg21}.}
This result follows from the shell-model equivalent of the Doering--Foias estimate for $D_1$ and the ordering in \eqref{eq:order}.
Unlike in the Navier--Stokes simulations \cite{donzis13}, the $\mathit{Re}^3$-scaling is now saturated for all $m$ (see figure~\ref{fig:Dm}, right panel; note, in the inset, that a damped oscillation is superposed to the $\Rey^3$ scaling).
Consequently, the separation between $\langle D_{1}\rangle_t $ and the time averages of the higher-order $D_m$ is not as dramatic as in the Navier--Stokes simulations.
We also observe that the time averages of $D_m$ get closer as $m$ increases; this behaviour is similar to that found for the three-dimensional Navier--Stokes equations. 

Finally, we conclude this section by establishing a relation between vorticity moments of different orders. To this end, it is convenient to introduce the quantities
\begin{equation}
\label{eq:Jm}
J_m = \sum_{j=1}^\infty k_j^{2m} \vert u_j\vert^{2m} = \Omega_m^{2m} = (\varpi_0^2 D_m)^m, \qquad m\geqslant 1\,.
\end{equation}
\begin{lemma} For $m-1\geqslant  p >  0$ and $q>0$,
\begin{equation}
D_{m}^{m(p+q)}\leqslant D_{m-p}^{q(m-p)} D_{m+q}^{p(m+q)}.
\label{eq:Dm-adjacent}
\end{equation}
\label{lemma}
\end{lemma}
\begin{proof}
Rewrite $J_m$ as
\begin{equation}
J_m = \sum_{j=1}^\infty k_j^{2m}\vert u_j\vert^{2m} = \sum_{j=1}^\infty  (k_j\vert u_j\vert)^{2\alpha} (k_j\vert u_j\vert)^{2\beta}
\end{equation}
with $\alpha+\beta=m$. A H\"older inequality gives
\begin{equation}
J_m \leqslant \bigg[\sum_{j=1}^\infty  (k_j\vert u_j\vert)^{2(m-p)}\bigg]^{\alpha/(m-p)}
\bigg[\sum_{j=1}^\infty  (k_j\vert u_j\vert)^{2(m+q)}\bigg]^{\beta/(m+q)} 
= J_{m-p}^{\alpha/(m-p)} J_{m+q}^{\beta/(m+q)} 
\label{eq:Jm-Holder}
\end{equation}
with 
\begin{equation}
\frac{\alpha}{m-p}+\frac{\beta}{m+q}=1
\end{equation}
and hence
\begin{equation}
\alpha=\frac{q(m-p)}{p+q}, \qquad \beta=\frac{p(m+q)}{p+q}\,.
\end{equation}
The inequality in \eqref{eq:Jm-Holder} becomes \eqref{eq:Dm-adjacent} upon expressing $J_m$  in terms of $D_m$.
\end{proof}

\section{Differential inequalities and absorbing balls for $D_m$}
\label{sect:shell-diff-ineq}

In this section, we derive the differential inequalities that describe the temporal evolution of the $D_m$.
\begin{proposition}
\label{prop:Dm}
For $1 < m < \infty$,
\begin{equation}
\label{eq:ineq-1}
\dfrac{\dot{D}_m}{2} \leqslant - \frac{\nu \varpi_0^2}{2E} \,D_m^2 \left(\frac{D_{m+1}}{D_m}\right)^{m+1} 
+c_0 \frac{E}{2\nu} D_m
+c_m \varpi_0  \Gr  D_m^{1/2},
\end{equation}
\REV{where
\begin{equation}
  c_0 = \Big(|a_1| \lambda^{-2} + |a_2| + |a_3| \lambda^2\Big)^2
  \label{eq:c0}
\end{equation}
and
\begin{equation}
  c_m = \bigg(\sum_{j=j_f}^{\infty}\lambda^{2m(j-j_f)}\vert\phi_j\vert^{2m} \bigg)^{1/2m}
  \label{eq:cm}
\end{equation}
is a function of the forcing shape alone.}
\end{proposition}
\begin{proof}
We first derive a differential equation for $J_m$ [see \eqref{eq:Jm} for the definition] and then convert into a differential equation for $D_m$.
To this end, we multiply \eqref{eq:shell} by $m\, k_j^{2m} (u_j^*)^{m}u_j^{m-1}$ and the complex conjugate of \eqref{eq:shell} by  $m\, k_j^{2m} u_j^{m}(u_j^*)^{m-1}$. We then add the two resulting equations and sum over $j$ to find:
\begin{subequations}
\label{eq:Jmdot}
\begin{eqnarray}
\dfrac{\dot{J}_m}{m}&=&
-2\nu \sum_{j=1}^\infty k_j^{2(m+1)}\vert u_j\vert^{2m} \label{eq:viscous}
\\
&&+\sum_{i=1}^\infty k_j^{2m} (u_j^*)^{m}u_j^{m-1} f_j + \mathrm{c.c.} \label{eq:forcing}
\\
&&+\mathrm{i} a_1\sum_{j=1}^\infty k_{j}^{2m} k_{j+1} (u_j^*)^m u_j^{m-1} u_{j+1}^* u_{j+2} + \mathrm{c.c.} \label{eq:nonlinear-a1}
\\
&&+\mathrm{i} a_2\sum_{j=1}^\infty k_{j}^{2m+1} (u_j^*)^m u_j^{m-1} u_{j-1}^* u_{j+1} + \mathrm{c.c.} \label{eq:nonlinear-a2}
\\
&&+\mathrm{i} a_3\sum_{j=1}^\infty k_{j}^{2m} k_{j-1} (u_j^*)^m u_j^{m-1} u_{j-1} u_{j-2} + \mathrm{c.c.}\,, \label{eq:nonlinear-a3}
\end{eqnarray}
\end{subequations}
where `c.c.' denotes the complex conjugate.

We start by estimating the viscous term in \eqref{eq:viscous}. By moving $\sup_{1\leqslant j} \vert u_j\vert^2$ out of the sum and using $\sup_{1\leqslant j} \vert u_j\vert^2\leqslant E(t)$, we get:
\begin{eqnarray}
\nonumber
J_{m+1}&=&\sum_{j=1}^\infty k_j^{2(m+1)}\vert u_j\vert^{2(m+1)} \leqslant
\sup_{1\leqslant j} \vert u_j\vert^2 \sum_{j=1}^\infty k_j^{2(m+1)}\vert u_j\vert^{2m}
\\
&\leqslant& E(t) \sum_{j=1}^\infty k_j^{2(m+1)}\vert u_j\vert^{2m} 
\end{eqnarray}
and hence
\begin{equation}
-2\nu\sum_{j=1}^\infty k_j^{2(m+1)}\vert u_j\vert^{2m} \leqslant -\frac{2\nu}{E}\, J_{m+1}\,.  \label{eq:estimate-viscous-m}
\end{equation}
We then move to the forcing term in \eqref{eq:forcing}. The triangle inequality and an H\"older's inequality give:
\begin{eqnarray}
\nonumber
\left\vert \sum_{j=1}^\infty k_j^{2m} (u_j^*)^{m}u_j^{m-1} f_j \right\vert  & \leqslant & \sum_{j=1}^\infty (k_j \vert u_j\vert)^{2m-1} (k_j\vert f_j\vert ) 
\\
\nonumber
&\leqslant& \bigg(\sum_{j=1}^\infty k_j^{2m} \vert u_j\vert^{2m}\bigg)^{\frac{2m-1}{2m}} \bigg(\sum_{j=1}^\infty k_j^{2m}\vert f_j\vert^{2m}\bigg)^{\frac{1}{2m}}
\\ 
&=& c_m \nu^2 k_f^4  \Gr  J_m^{\frac{2m-1}{2m}},  \label{eq:estimate-forcing-m}
\end{eqnarray}
where
\begin{equation}
  \REV{c_m = \lambda^{-j_f}\bigg(\sum_{j=j_f}^{\infty}\lambda^{2jm}\vert\phi_j\vert^{2m} \bigg)^{1/2m}}
\end{equation}
\REV{and we have used $\sum_{j=1}^\infty k_j^{2m}\vert\phi_j\vert^{2m} \leqslant \big(\sum_{j=1}^\infty k_j^{2}\vert\phi_j\vert^{2}\big)^{m}<\infty$.}
We are left to estimate the nonlinear terms in \eqref{eq:nonlinear-a1} to \eqref{eq:nonlinear-a3}. We only derive the estimate of the term with cofficient $a_1$ in \eqref{eq:nonlinear-a1}; the estimates of the other two terms follow in the same manner. By using $k_{j+2}=\lambda^{-2}k_j$, a H\"older inequality, and the Cauchy--Schwarz inequality, we find:
\begin{align}
\nonumber
\left\vert\sum_{j=1}^\infty k_{j}^{2m} k_{j+1} (u_j^*)^m u_j^{m-1} u_{j+1}^* u_{j+2}\right\vert &
\leqslant \lambda^{-2} \sum_{j=1}^\infty \left(k_{j} \vert u_j\vert \right)^{2m-1} \left(k_{j+1} \vert u_{j+1}\vert\right) \left(k_{j+2} \vert u_{j+2}\vert\right)
\\
\nonumber
&\leqslant \lambda^{-2} \sum_{j=1}^\infty k_j^{2m+1} \vert u_j\vert^{2m+1}  
\\
\nonumber
&= \lambda^{-2} \sum_{j=1}^\infty (k_j \vert u_j\vert)^{m+1} (k_j \vert u_j\vert)^{m}
\\
&\leqslant  \lambda^{-2}  J_{m+1}^{1/2}J_m^{1/2}. \label{eq:estimate-nonlinear-m}
\end{align}
By proceedings in the same manner for the other two nonlinear terms, we find
\begin{equation}
\vert\text{nonlinear terms}\vert \leqslant 2c_0^{1/2} J_{m+1}^{1/2}J_m^{1/2}.
\end{equation}
\REV{where $c_0^{1/2} = \vert a_1\vert \lambda^{-2} + \vert a_2\vert + \vert a_3\vert \lambda^2$.}
By combining  \eqref{eq:estimate-viscous-m},  \eqref{eq:estimate-forcing-m}, and  \eqref{eq:estimate-nonlinear-m}, we find:
\begin{equation}
\label{eq:ineq-Jm}
\dfrac{\dot{J}_m}{2m}\leqslant
-\frac{\nu}{E}\, J_{m+1} +c_0^{1/2} J_{m+1}^{1/2} J_{m}^{1/2} + c_m \varpi_0^2 \Gr  J_m^{\frac{2m-1}{2m}}.
\end{equation}
We now apply Young's inequality to the second term on the right-hand side to get:
\begin{equation}
\dfrac{\dot{J}_m}{2m}\leqslant
-\frac{\nu}{2E}\, J_{m+1} +c_0 \frac{E}{2\nu} J_{m} + c_m \varpi_0^2 \Gr  J_m^{\frac{2m-1}{2m}}.
\end{equation}
This inequality becomes \eqref{eq:ineq-1} upon substituting for $D_m$ from \eqref{eq:Jm}.
\end{proof}


Note that, in the proof of Proposition~\ref{prop:Dm}, the estimates of the nonlinear and forcing terms parallel those of the corresponding terms in the Navier--Stokes equations \cite{gibbon14,gibbon2016ima}. It is the estimate of the viscous term that differs. We have indeed used  $\sup_{1\leqslant j} \vert u_j\vert^2 \leqslant E(t)$, which is specific to shell models.




The inequality in \eqref{eq:ineq-1} implies that $D_m$ are bounded. This can be shown by ignoring the negative term on the right-hand side of \eqref{eq:ineq-1} and using the point-wise estimate on $E$ in \eqref{eq:point-wise-energy}. However, such bound would be exponentially growing in time.
We will show below that in fact there exists an absorbing ball for $D_m$ for all $m$. To this end, we first write a modified version of the differential inequality for $D_m$ that brings in the ratio of $D_m$ and $D_1$.
\begin{corollary}
For $1 < m < \infty$,
\begin{equation}
\dfrac{\dot{D}_m}{2} \leqslant - \frac{\nu \varpi_0^2}{2E} \, D_m^2 \left(\frac{D_m}{D_1}\right)^{\frac{1}{m-1}}
+c_0\, \frac{E}{2\nu} D_m
+c_m \varpi_0  \Gr  D_m^{1/2},
\label{eq:ineq-2}
\end{equation}
\REV{where $c_0$ and $c_m$ are defined in \eqref{eq:c0} and \eqref{eq:cm}, respectively.}
\end{corollary}
\begin{proof}
For $p=m-1$ and $q=1$, Lemma~\ref{lemma} yields
\begin{equation}
D_m^{m^2} \leqslant D_1 D_{m+1}^{m^2-1}.
\label{eq:three}
\end{equation}
The result follows upon using \eqref{eq:three} to bound the ratio $D_{m+1}/D_m$ in \eqref{eq:ineq-1}.
\end{proof}
We then derive a differential inequality for $D_1$. For $m=1$, the viscous term can indeed be estimated differently (in analogy with the estimate for the Navier--Stokes equations \cite{gibbon14,gibbon2016ima}).
\begin{proposition}
\label{prop:D1}
$D_1$ satisfies the differential inequality
\begin{equation}
\label{eq:ineq-D1}
\dfrac{\dot{D}_1}{2\varpi_0}
\leqslant -\frac{\nu\varpi_0}{E}\, D_{1}^2 +c_0 D_{1}^{3/2} + c_1 \Gr  D_1^{1/2},
\end{equation}
\REV{where $c_0$ and $c_1$ are defined in \eqref{eq:c0} and \eqref{eq:cm}, respectively.}
\end{proposition}
\begin{proof}
The evolution equation for $J_1$ can be deduced from \eqref{eq:Jmdot} by setting  $m=1$. The forcing and nonlinear terms can be estimated similarly to the case $m>1$. However, the estimate of the viscous term differs.
By using the Cauchy--Schwarz inequality, we indeed get
\begin{equation}
\sum_{j=1}^\infty k_j^2 \vert u_j\vert^2 = \sum_{j=1}^\infty (k_j^2 \vert u_j\vert) \vert u_j\vert \leqslant E^{1/2} \bigg(\sum_{j=1}^\infty k_j^4 \vert u_j\vert^2\bigg)^{1/2}, 
\end{equation}
whence
\begin{equation}
\label{eq:m1}
-2\nu\sum_{j=1}^\infty k_j^{4}\vert u_j\vert^{2} \leqslant -\frac{2\nu}{E} J_1^2.
\end{equation}
Therefore,  the differential inequality for $J_1$ becomes
\begin{equation}
\dfrac{\dot{J}_1}{2}\leqslant
-\frac{\nu}{E}\, J_{1}^2 +c_0 J_{2}^{1/2} J_{1}^{1/2} + c_1 \varpi_0^2 \Gr  J_1^{1/2},
\label{eq:J1}
\end{equation}
which upon substituting for $D_1$, gives
\begin{equation}
\dfrac{\dot{D}_1}{2}\leqslant
-\frac{\nu\varpi_0^2}{E}\, D_{1}^2 +c_0 \varpi_0 D_{2} D_{1}^{1/2} + c_1 \varpi_0 \Gr  D_1^{1/2}.
\label{eq:d2d1}
\end{equation}
The result in \eqref{eq:ineq-D1} follows from using $D_2\leqslant D_1$ in the latter inequality.
\end{proof}
In the proof of Proposition~\ref{prop:D1}, the bound in \eqref{eq:m1} has a Navier--Stokes analogue, while the step that uses $D_2\leqslant D_1$ is specific to shell models. The bound $D_2\leqslant D_1$ produces the $D_1^{3/2}$ term on the right-hand side, which should be contrasted with the $D_1^3$ term  in the Navier--Stokes equations.
This difference is what yields point-wise control over $D_1$ (and ultimately over all $D_m$) in shell models, since for large values of $D_1$ the viscous term is able to counter the nonlinear term. The boundedness of $D_1$ can then be used to prove explicit point-wise upper bounds for $D_m$.
\REV{
\begin{corollary}
For $1\leqslant m < \infty$, there exists an absorbing ball for $D_m$ such that
\begin{equation}
\label{eq:point-wise-D1}
\varlimsup_{t\to\infty} D_1(t) \leqslant c_1^{2/3} \rho_f^{8/3} \Gr ^2 + c_0^2 \,\rho_f^8 \Gr ^4
\end{equation}
and, for $1<m<\infty$,
\begin{equation}
\label{eq:point-wise-Dm}
\varlimsup_{t\to\infty} D_m(t) \leqslant 
\begin{cases}
(2c_m)^\frac{2(m-1)}{3m-1} c_1^\frac{4}{3(3m-1)} \rho_f^{8/3} \Gr^2, & (\Gr\ll 1),
\\[2mm]
c_0^\frac{m+1}{m} \rho_f^8 \Gr^4, & (\Gr\gg 1),
\end{cases}
\end{equation}
where $c_0$ and $c_m$ are defined in \eqref{eq:c0} and \eqref{eq:cm}, respectively.
\end{corollary}
}
\begin{proof}
For $m=1$, we  apply a standard comparison theorem for ordinary differential equations to \eqref{eq:ineq-D1}  (see Appendix~B in Ref.~\cite{gp07})
and make use of the point-wise bound on the total energy in \eqref{eq:point-wise-energy}. 
For $m>1$, \eqref{eq:ineq-2} together with the bounds in \eqref{eq:point-wise-energy} and \eqref{eq:point-wise-D1} yields \eqref{eq:point-wise-Dm}.
\end{proof}
The point-wise bound for $D_1$ was also obtained in Ref.~\cite{vg21}, but in a different way; the bound for $m>1$ is new.

\section{Discussion}
\label{sect:discussion}

The differential inequalities for the rescaled vorticity moments display a crucial difference in the Navier--Stokes equations and the shell model.
The contribution coming from the nonlinear term is $D_m^3$ in the Navier--Stokes equations, while it is $D_m^{3/2}$,  and hence significantly weaker, in the shell model. As a consequence, for the Navier--Stokes equations, $D_m$ is known to be bounded only for small initial data or for short times; for the shell model, there exist absorbing balls for all $D_m$.
\REV{This difference is due to  the fact the shell model essentially is a {\em scalar} representation of the Navier--Stokes equations in Fourier space, whereas the nonlinear term in the Navier--Stokes equations results from the interaction of triads of three-dimensional wavevectors. Certain simplifications that hold for the shell model are therefore not allowed for the Navier--Stokes equations. A way to appreciate this fact is to consider the derivation of the differential inequality for $D_1$.  Recall the definition of $D_1$ for the three-dimensional Navier--Stokes equations in \eqref{Dmdef}. The Navier--Stokes counterpart of \eqref{eq:d2d1} is (see, e.g., Ref.~\cite{gibbon14})
\begin{equation}
\frac{\dot{D}_1}{2\varpi_0} \leqslant  - \frac{\nu^2 L}{4E} D_1^2 +  \frac{L^{5/2}}{\nu^2} \, D_1^{1/2} \Vert\bm\omega\Vert_4^2 + \mathit{Gr} D_1^{1/2}.
\end{equation}
Now, Parseval's theorem implies that $D_1=\varpi_0^{-2}\sum_{\bm k\neq 0} k^2\vert\hat{\bm u}_{\bm k}\vert^2$. Therefore, $D_1$ has the same structure in the Navier--Stokes equations and in the shell model. However, in general, $\Vert\bm\omega\Vert_4^4$ is not related to $\sum_{\bm k\neq 0} k^4\vert\hat{\bm u}_{\bm k}\vert^4$ in a simple manner. Consequently, the steps that, in the shell model, lead to a subdominant nonlinear term cannot be repeated for the Navier--Stokes equations, and the nonlinear contribution to the differential inequality for $D_1$ can only be estimated as $D_1^3$, as in \eqref{eq:ineq-D1-NS}.
}

\REV{If $\Gr \ll 1$,  \eqref{eq:point-wise-D1} and \eqref{eq:point-wise-Dm} indicate that $D_m$ scales as $\Gr^2$ for all $m\geqslant 1$. Contrastingly, at large $\Gr$, the forcing contributions are subdominant, and the bound scales as $\Gr ^4$ uniformly in $m$. }
While this fact remains consistent with the ordering $D_{m+1}\leqslant D_m$, it shows that the values of $D_m$ for different $m$ can get very close to each other during the time evolution.
By using \eqref{eq:Gr-Re_shell} in \eqref{eq:point-wise-Dm}, it is also possible to see that, for large $\Gr $, 
\begin{equation}
\varlimsup_{t\to\infty} D_m(t) \leqslant \tilde{c}\, \Rey ^8,
\end{equation}
\REV{where $\tilde{c}=c''\delta_4$.}
Therefore, $D_m$ can display very large escursions from its time average, which only scales as $\Rey ^3$. This fact is consistent with the spiky behaviour of the time series of $D_m$ that is typically observed in numerical simulations of shell models (figure~\ref{fig:Dm}).

\begin{figure}[t]
\includegraphics[width=0.5\textwidth]{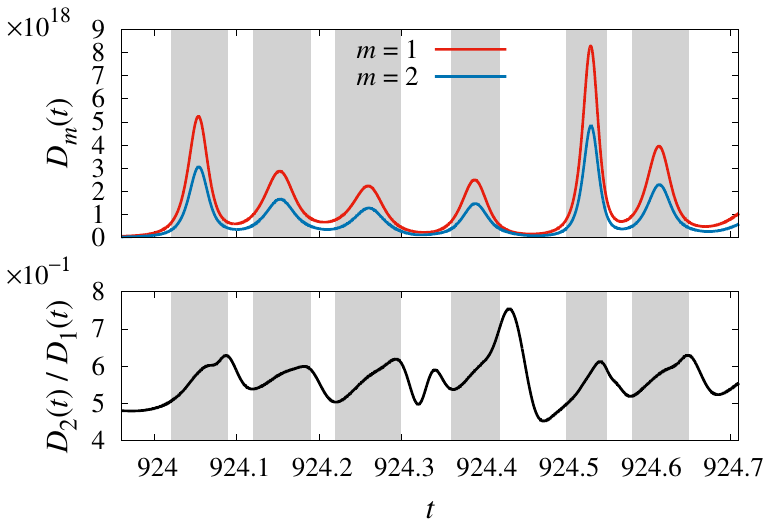}%
\hfill%
\includegraphics[width=0.5\textwidth]{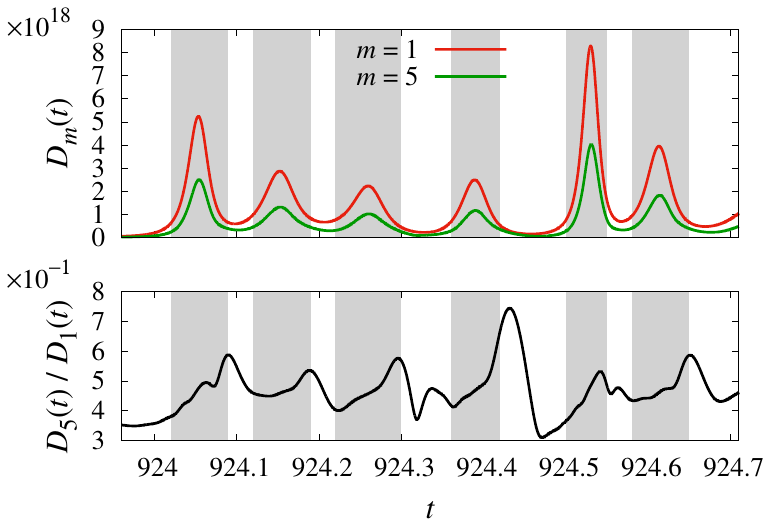}%
\caption{Representative time series of $D_1$, $D_m$, and $D_m/D_1$  for $m=2$ and $m=5$ and $\Rey \approx 7.4\times 10^5$. The time series  shows that \REV{ the ratio $D_m/D_1$ grows during a spike and reaches a maximum at the end of the spike.} The details of the simulation are given in the caption of figure~\ref{fig:Dm}.}
\label{fig:Dm/D1}
\end{figure}

The differential inequalities \eqref{eq:ineq-2} and \eqref{eq:ineq-D1} indicate that $D_1$ `drives' the temporal evolution of all $D_m$. The time evolution of $D_1$ is indeed independent of that of higher-order $D_m$. When $D_1$ is sufficiently small the nonlinear term dominates the right-hand side of \eqref{eq:ineq-D1} and $D_1$ can grow rapidly, in principle as fast as $(t-t_\star)^{-2}$  for some $t_\star>0$.
The growth of $D_1$ ends when the viscous term becomes large enough to prevail over the nonlinear term, at which point $D_1$ starts decreasing. The temporal evolution of $D_1$ is therefore an alternation of quiescent periods and sudden spikes. 
When $D_1$ is in a quiescent period the higher-order $D_m$ must also take small values because $D_m\leqslant D_1$ for all $m$. However, when $D_1$ increases the viscous term in \eqref{eq:ineq-2} further loses power, and the higher-order $D_m$ can grow (at most exponentially).
The growth of $D_m$ continues until the viscous term gains enough strength. Figure \ref{fig:Dm/D1} shows that, for $m>1$, this happens as a result of two effects: 1) the exponent of $D_m$ is higher in the viscous term than in the nonlinear term and \REV{2) the ratio $D_m/D_1$ grows during a spike. The latter effect makes the factor $(D_m/D_1)^{1/(m-1)}$ closer to unity and hence  gives additional strength to the dissipation term. The ratio $D_m/D_1$ reaches a maximum at the end of the spike.}

\begin{figure}[t]
\centering
\includegraphics[width=0.5\textwidth]{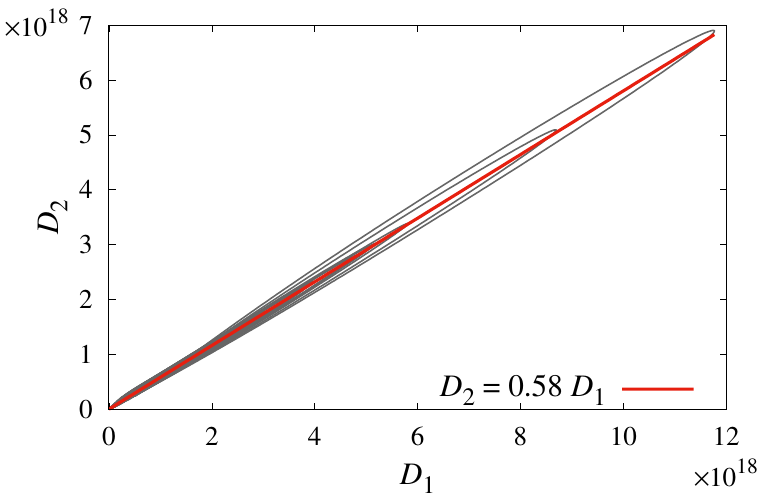}%
\hfill
\includegraphics[width=0.5\textwidth]{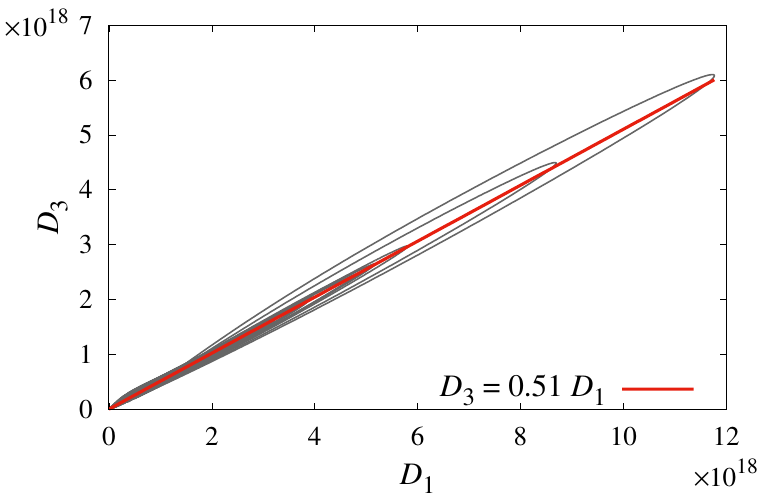}%
\\
\includegraphics[width=0.5\textwidth]{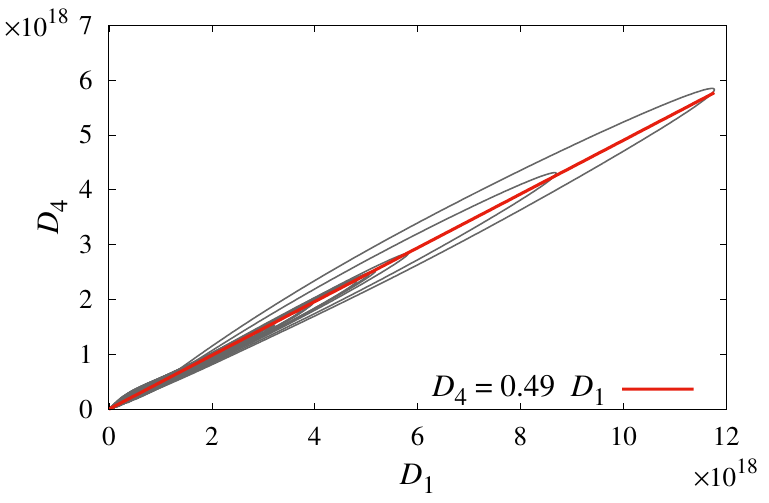}%
\hfill
\includegraphics[width=0.5\textwidth]{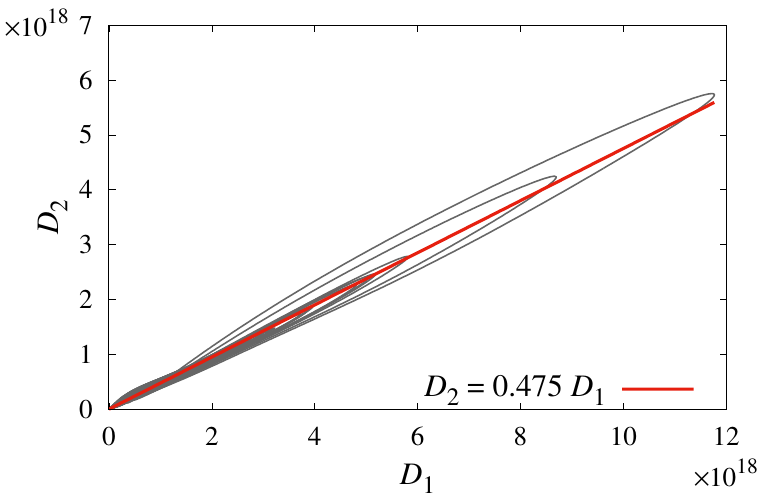}%
\caption{$D_m$ versus $D_1$ for $m=2,\dots,5$ and $\Rey \approx 7.4\times 10^5$.
The details of the simulation are given in the caption of figure~\ref{fig:Dm}.}
\label{fig:Dm-vs-D1}
\end{figure}

Comparing \eqref{eq:ineq-D1} with \eqref{eq:Dm-lambda},  \REV{as well as \eqref{eq:ns-pointwise} with \eqref{eq:point-wise-D1}}, indicates that the shell model corresponds to $\mu_m=1$. This value of $\mu_m$ lies at the bottom of the range for which $D_1$ comes under control in the Navier--Stokes equations and indeed corresponds to the maximum depletion of nonlinearity that is achievable according to the high-low frequency slaving argument.
In fact, the value $\mu_m=1$ can  be obtained by suitably adapting the high-low frequency slaving approach to shell models. Since for shell models $\alpha_m=2$, inequality \eqref{eq:condition-lambda} together with the constraint $\mu_m\geqslant 1$ gives $\mu_m=1$ for all $m$ and all times. Therefore, the $D_m$ are expected to be connected with $D_1$ via a linear relation such as \REV{$D_m = b_m D_1$ with $b_m$ a dimensionless constant}.
Figure \ref{fig:Dm-vs-D1} shows that in numerical simulations $D_m$ and $D_1$ indeed tend to cluster around a straight line, with a slope that varies weakly with $m$. \REV{The time series in figure~\ref{fig:Dm/D1} also indicate that the ratios $D_m/D_1$ oscillate around a constant value.}

Reference~\cite{vg21} established a formal connection between shell models and the Navier--Stokes equations `on a point'. The estimates of the Sobolev norms of the velocity for the $d$-dimensional Navier--Stokes equations indeed reduce to their shell-model analogues in the limit $d\to 0$. That correspondence carries over to the results of this paper. Let us denote, for $1\leqslant m\leqslant\infty$ and $d=1,2,3$, 
\begin{equation}
\Omega_{m,d}(t)=L^{-d/2m} \Vert\bom\Vert_{L^{2m}([0,L]^d)}
\end{equation}
and
\begin{equation}
D_{m,d}=\big(\varpi_0^{-1}\Omega_m\big)^{(4-d)\alpha_{1,m,d}}, \qquad \alpha_{1,m,d}=\dfrac{2m}{4m-d}\,.
\end{equation}
For the $d$-dimensional Navier--Stokes equations ($d=2,3)$ and the Burgers equation ($d=1$), we have estimates of the form \cite{gibbon2020epl,gibbon2023}
\begin{equation}
\left\langle D_{m,d}\right\rangle_t \leqslant c_{1,m,d} \Rey^3 + O(t^{-1})\,.
\label{eq:Dm-d}
\end{equation}
In addition, the differential inequality for $m=1$ becomes \cite{gibbon14}
\begin{equation}
\label{eq:NS-d}
\frac{\dot{D}_{1,d} }{2\varpi_0}\leqslant -\left(\frac{4-d}{4}\right)\frac{\nu^2 L}{E} \, D_{1,d}^2+ \tilde{c}_d D_{1,d}^{\frac{6-d}{4-d}} + \Gr D_{1,d}^{1/2}
\end{equation}
\REV{with $\tilde{c}_d$ a dimensionless constant.}
It can be observed that the vortex-stretching term `weakens' with decreasing $d$ and is balanced by the viscous term for $d=2$, which is a critical dimension for regularity.
Incidentally, this fact is connected with the formation of low-dimensional structures of intense strain and vorticity in turbulent flows~\cite{gibbon2020epl}.
It is clear that \eqref{eq:NS-d} reduces to the differential inequality for $D_1$ in the shell model when $d\to 0$.

Furthermore, the high--low frequency scaling argument of \S\ref{sect:high-low} can be extended as such to the $d$-dimensional Navier--Stokes equations. The only difference is in the range of variation of $\mu_m(\tau)$. If it is indeed required that $\left\langle F_m^{(4-d)\alpha_{1,m,d}}\right\rangle_t$ can be estimated in terms of $\big\langle F_m^2\big\rangle_t$ as in \eqref{eq:Dm-d} [note that $(4-d)\alpha_{1,1,d}=2$ for all $d$], then $\mu_{m}(\tau)$ must satisfy
\begin{equation}
1 \leqslant  \mu_{m}(\tau) \leqslant \dfrac{4}{4-d}\,.
\end{equation}
Once again, we observe that formally decreasing $d$ shrinks the range of variation of $\mu_m(\tau)$ towards its lower bound and therefore goes in the direction of more regularity, with full regularity being achieved for $d=2$. Furthermore, the shell model value $\mu_m(\tau)=1$ corresponds to the limit $d\to 0$. 

We also note that the derivations in \S\ref{sect:shell-diff-ineq} are rigorous, because the solutions of the shell model are regular.
However, inequality \eqref{eq:ineq-D1} can be used for an alternative proof of regularity. Such a proof is based on a standard contradiction strategy that assumes a maximal time interval of regularity, $[0,T^*]$, and then proves a contradiction in the limit $t\to T^*$.

\REV{Moreover, the results in \S~\ref{sect:shell-diff-ineq} hold irrespective of the values of the cofficients of the shell model, the only condition being that $a_1$, $a_2$, $a_3$ sum up to zero to preserve energy.
  Therefore, our study covers both the three- and the two-dimensional regimes of the shell model \cite{shell-inverse}.}

\REV{A possible extension of this study would be to compare the moments of the Els\"asser vorticity field in magnetohydrodynamic turbulence \cite{gibbonMHD} with their analogues in shell models of magnetohydrodynamics \cite{psf13,bsdp98}, so as to establish a rigorous connection between the two systems.}
\vskip6pt

\enlargethispage{20pt}

The authors would like to thank the Isaac Newton Institute for Mathematical Sciences, Cambridge, for support and hospitality during the programme ``Anti-diffusive dynamics\,: from sub-cellular to astrophysical scales'', where work on this paper was undertaken. This work was supported by EPSRC grant no EP/R014604/1. D.V. acknowledges his Associateship with the International Centre for Theoretical Sciences, Tata Institute of Fundamental Research, Bangalore, India and thanks Ritwik Mukherjee for a careful reading of the manuscript.



\begin{thebibliography}{99}

\bibitem{frisch1995}
Frisch U. 1995 {\em Turbulence: The Legacy of A.~N.~Kolmogorov}.
Cambridge: Cambridge University Press.

\bibitem{bjpv98}
Bohr T, Jensen MH, Paladin G, Vulpiani A. 1998 {\em Dynamical Systems Approach
  to Turbulence}.
Cambridge: Cambridge University Press.

\bibitem{biferale2003}
Biferale L. 2003  Shell models of energy cascade in turbulence. {\em Annu. Rev.
  Fluid Mech.} \textbf{35}, 441--468.

\bibitem{psf13}
Plunian F, Stepanov R, Frick P. 2013  Shell models of magnetohydrodynamic
  turbulence. {\em Phys. Rep.} \textbf{523}, 1--60.

\bibitem{bt23}
Benzi R, Toschi F. 2023 Lectures on turbulence. {\em Phys. Rep.} \textbf{1021}, 1--106.

\bibitem{cdf23}
Cheskidov A, Dai M, Friedlander S. 2023  Dyadic models for fluid equations: A
  survey. {\em J. Math. Fluid Mech.} \textbf{25}, 62.

\bibitem{dewit2024}
de~Wit XM, Ortali G, Corbetta A, Mailybaev AA, Biferale L, Toschi F. 2024
  Extreme statistics and extreme events in dynamical models of turbulence. {\em
  Phys.~Rev. {\rm E}} \textbf{109}, 055106.

\bibitem{clt06}
Constantin P, Levant B, Titi ES. 2006  Analytic study of shell models of
  turbulence. {\em Physica {\rm D}} \textbf{219}, 120--141.

\bibitem{clt07a}
Constantin P, Levant B, Titi ES. 2007a  Sharp lower bounds for the dimension of
  the global attractor of the {Sabra} shell model of turbulence. {\em J. Stat.
  Phys.} \textbf{127}, 1173--1192.

\bibitem{clt07b}
Constantin P, Levant B, Titi ES. 2007b  Regularity of inviscid shell models of
  turbulence. {\em Phys. Rev. {\rm E}} \textbf{75}, 016304.

\bibitem{mailybaev12}
Mailybaev AA. 2012  Renormalization and universality of blowup in hydrodynamic
  flows. {\em Phys. Rev. {\rm E}} \textbf{85}, 066317.

\bibitem{bbbf06}
Barbato D, Barsanti M, Bessaih H, Flandoli F. 2006  Some rigorous results on a
  stochastic GOY model. {\em J. Stat. Phys.} \textbf{125}, 677--716.

\bibitem{bf12}
Bessaih H, Ferrario B. 2012  Invariant Gibbs measures of the energy for shell
  models of turbulence: the inviscid and viscous cases. {\em Nonlinearity}
  \textbf{25}, 1075--1097.

\bibitem{bgas16}
Bessaih H, Garrido-Atienza MJ, Schmalfuss B. 2016a  Stochastic shell models
  driven by a multiplicative fractional Brownian-motion. {\em Physica {\rm D}}
  \textbf{320}, 38--56.

\bibitem{bhr16}
Bessaih H, Hausenblas E, Razafimandimby PA. 2016b  Ergodicity of stochastic
  shell models driven by pure jump noise. {\em SIAM J. Math. Anal.}
  \textbf{48}, 1423--1458.

\bibitem{vg21}
Vincenzi D, Gibbon JD. 2021  How close are shell models to the 3{D}
  {N}avier--{S}tokes equations?. {\em Nonlinearity} \textbf{34}, 5821--5843.

\bibitem{dg95}
Doering CR, Gibbon JD. 1995 {\em Applied Analysis of the {N}avier--{S}tokes
  Equations}.
Cambridge: Cambridge University Press.

\bibitem{foias2001}
Foias C, Manley O, Rosa R, Temam R. 2001 {\em Navier--Stokes Equations and
  Turbulence}.
Cambridge: Cambridge University Press.

\bibitem{ld08}
Lu L, Doering CR. 2008  Limits on Enstrophy Growth for Solutions of the
  Three-dimensional {Navier--Stokes} equations. {\em Indiana Univ. J. Math.}
  \textbf{57}, 2693--2727.

\bibitem{doering2009}
Doering CR. 2009  The 3{D} {N}avier--{S}tokes problem. {\em Annu. Rev. Fluid
  Mech.} \textbf{41}, 109--128.

\bibitem{robinson2016}
Robinson JC, Rodrigo JL, Sadowski W. 2016 {\em The Three-Dimensional
  Navier--Stokes Equations}.
Cambridge: Cambridge University Press.

\bibitem{robinson2020}
Robinson JC. 2020  The {Navier--Stokes} regularity problem. {\em Phil. Trans.
  R. Soc. {\rm A}} \textbf{378}, 20190526.

\bibitem{vicol2022}
Bedrossian J, Vicol V. 2022 {\em The Mathematical Analysis of the
  Incompressible Euler and Navier--Stokes Equations}.
Providence, RI: American Mathematical Society.

\bibitem{Leray1934}
Leray. J. 1934  Sur le mouvement d'un liquide visqueux emplissant l'espace.
  {\em Acta Math.} \textbf{63}, 193--248.

\bibitem{fgt1981}
Foias C, Guillop\'e C, Temam R. 1981  New a priori estimates for the
  {N}avier--{S}tokes equations in dimension 3. {\em Commun. Partial Diff. Equ.}
  \textbf{6}, 329--359.

\bibitem{gibbon19}
Gibbon JD. 2019  Weak and strong solutions of the {3$D$} {Navier--Stokes}
  equations and their relation to a chessboard of convergent inverse length
  scales. {\em J. Nonlin. Sci.} \textbf{29}, 215--228.

\bibitem{gibbon2020epl}
Gibbon JD. 2020  Intermittency, cascades and thin sets in three-dimensional
  {Navier--Stokes} turbulence. {\em Europhys. Lett.} \textbf{131}, 64001.

\bibitem{gibbon2023}
Gibbon JD. 2023  Identifying the multifractal set on which energy dissipates in
  a turbulent {Navier--Stokes} fluid. {\em Physica {\rm D}} \textbf{445},
  133654.

\bibitem{gibbon12cms}
Gibbon JD. 2012  A hierarchy of length scales for weak solutions of the
  three-dimensional {Navier--Stokes} equations. {\em Commun. Math. Sci.}
  \textbf{10}, 131--136.

\bibitem{donzis13}
Donzis D, Gibbon JD, Gupta A, Kerr RM, Pandit R, Vincenzi D. 2013  Vorticity
  moments in four numerical simulations of the {3D} {Navier--Stokes} equations.
  {\em J. Fluid Mech.} \textbf{732}, 316--331.

\bibitem{kerr2013}
Kerr RM. 2013 Bounds for Euler from vorticity moments and line divergence. {\em J. Fluid Mech.}
\textbf{729}, R2.

\bibitem{df02}
Doering CR, Foias C. 2002  Energy dissipation in body-forced turbulence. {\em
  J. Fluid Mech.} \textbf{467}, 289--306.

\bibitem{gibbon14}
Gibbon JD, Donzis D, Gupta A, Kerr RM, Pandit R, Vincenzi D. 2014  Regimes of
  nonlinear depletion and regularity in the {3D Navier--Stokes} equations. {\em
  Nonlinearity} \textbf{27}, 2605--2625.

\bibitem{gibbon2016ima}
Gibbon JD. 2016  High–low frequency slaving and regularity issues in the {3D}
  {Navier--Stokes} equations. {\em IMA J. Appl. Math.} \textbf{81}, 308--320.

\bibitem{fst88}
Foias C, Sell GR, Temam R. 1988  Inertial manifolds for nonlinear evolutionary
  equations. {\em J. Diff. Equ.} \textbf{73}, 309--353.

\bibitem{ft91}
Foias C, Titi ES. 1991  Determining nodes, finite difference schemes and
  inertial manifolds. {\em Nonlinearity} \textbf{4}, 135--153.

\bibitem{foias1988pla}
Foias C, Jolly MS, Kevrekidis IG, Sell GR, Titi ES. 1988  On the computation of
  inertial manifolds. {\em Phys. Lett. {\rm A}} \textbf{131}, 433--436.

\bibitem{sabra}
L'vov VS, Podivilov E, Pomyalov A, Procaccia I, Vandembroucq D. 1998  Improved
  shell model of turbulence. {\em Phys. Rev. {\rm E}} \textbf{58}, 1811--1822.

\bibitem{g73}
Gledzer EB. 1973  System of hydrodynamic type admitting two quadratic integrals
  of motion. {\em Sov. Phys. Dokl.} \textbf{18}, 216--217.

\bibitem{yo87}
Yamada M, Ohkitani K. 1987  Lyapunov spectrum of a chaotic model of
  three-dimensional turbulence. {\em J. Phys. Soc. Japan} \textbf{56},
  4210--4213.

\bibitem{pisarenko1993}
Pisarenko D, Biferale L, Courvoisier D, Frisch U, Vergassola M. 1993  Further
  results on multifractality in shell models. {\em Phys. Fluids} \textbf{5},
  2533--2538.

\bibitem{gp07}
Gibbon JD, Pavliotis GA. 2007  Estimates for the two-dimensional
{N}avier--{S}tokes equations in terms of the {R}eynolds number. {\em J. Math. Phys.} \textbf{48}, 065202.

\bibitem{shell-inverse}
  Gilbert T, L'vov VS, Pomyalov A, Procaccia I. 2002
  Inverse cascade regime in shell models of two-dimensional turbulence.
  {\em Phys.~Rev.~Lett.}~\textbf{89}, 074501.
  

\bibitem{gibbonMHD}
Gibbon JD, Gupta A, Krstulovic G, Pandit R, Politano H, Ponty Y, Pouquet A, Sahoo G, Stawarz J. 2016
Depletion of nonlinearity in magnetohydrodynamic turbulence: Insights from analysis and simulations.
{\em Phys.~Rev.~{\rm E}} \textbf{93}, 043104.

\bibitem{bsdp98}
Basu A, Sain A, Dhar SK, Pandit R. 1998
Multiscaling in models of magnetohydrodynamic turbulence.
{\em Phys.~Rev.~Lett.}~\textbf{81}, 2687--2690.

\end{thebibliography}
\end{document}